\documentclass[letter]{aa}

\usepackage{natbib}
\usepackage{graphicx}
\usepackage{txfonts}
\usepackage{color}

\newcommand{\msun}{$M_{\odot}$}

\begin{document}

\title{Spirals in protoplanetary disks from photon travel time}

\author{
	M. Kama\inst{1}
	\and
	P. Pinilla\inst{1}
	\and
	A. N. Heays\inst{1}
}

\institute{Leiden Observatory, P.O. Box 9513, 2300 RA, Leiden, The Netherlands}

\date{}

\abstract{
Spiral structures are a common feature in scattered-light images of protoplanetary disks, and of great interest as possible tracers of the presence of planets. However, other mechanisms have been put foward to explain them, including self-gravity, disk-envelope interactions, and dead zone boundaries. These mechanisms explain many spirals very well, but are unable to easily account for very loosely wound spirals and single spiral arms. We study the effect of light travel time on the shape of a shadow cast by a clump orbiting close (within ${\sim}1\,$au) of the central star, where there can be significant orbital motion during the light travel time from the clump to the outer disk and then to the sky plane. This delay in light rays reaching the sky plane gives rise to a variety of spiral- and arc-shaped shadows, which we describe with a general fitting formula for a flared, inclined disk. 
}
   \keywords{Scattering / Methods: observational / Protoplanetary disks}
   \maketitle

\section{Introduction}

Spiral arms are often seen in scattered light images of protoplanetary disks \citep[e.g.][]{Gradyetal2001, Hashimotoetal2011, Boccalettietal2013, Gradyetal2013, Avenhausetal2014, Benistyetal2015, Wagneretal2015, Stolkeretal2016} and there is much interest in understanding their origins -- whether due to perturbations by giant planets, disk-envelope interactions, instabilities, or other phenomena. We present a new mechanism for creating scattered light spirals: the finite speed of light can cause a shadow from the inner disk to morph into a spiral, with properties determined by the disk and viewing geometry, and the orbital motion of the shadowing clump. We discuss observational evidence for spiral shadows, and provide a fitting formula for the general case of a flared, inclined disk.

A common hypothesis for the origin of spiral pressure waves in disks is that they are driven by giant planets orbiting in the disk \citep{GoldreichTremaine1979, Rafikov2002, Mutoetal2012, Dongetal2015}, although a good agreement between current planet-disk interaction modelling and observations is still under debate \citep[e.g.][]{Juhaszetal2015}. Alternative explanations for the origin of the spiral arms include gravitational instabilities when the disk is sufficiently cold and massive \citep[e.g.][]{LodatoRice2004}, shocks between a disk and a potential inflow from an external envelope \citep{Lesuretal2015}, or turbulent waves propagating into dead zones \citep{Lyraetal2015}. 

Some disks display a time-variable scattered light brightness, which is considered evidence for shadowing by inner disk structures which have a short orbital time \citep[e.g. HD~163296 and J160421,][]{Garufietal2014, Pinillaetal2015}. Such variable extinction in the inner disk also explains several types of variation in high-precision and -cadence lightcurves of highly inclined disk-hosting stars such as AA~Tau \citep{Bouvieretal2007} and a number of sources in the NGC~2264 cluster \citep{McGinnisetal2015, Staufferetal2015}. In particular, there is a class of sources with long-lived, regular, and narrow dips in their lightcurves. The inner disk features causing such dimmings could cast azimuthally narrow shadows across the outer disk. 

The finite speed of light leads to the propagation of light echoes from time-variable sources such as supernovae and outbursting stars. Light echoes on a scale of $10\,000\,$au have been resolved around young stars \citep{Ortizetal2010}. Protoplanetary disks themselves, with scattered-light radii of ${\sim}100$ to $1000\,$au \citep{Burrowsetal1996, McCaughreanOdell1996, Padgettetal1999, McCaughreanetal1998, Gradyetal2001, Wisniewskietal2008}, are large enough for travel time to play a role in the propagation of light from the central star. Spatially unresolved light travel time effects in a disk on ${\sim}100\,$au scales were identified in a multi-wavelength time-resolved but spatially unresolved study by \citet{Mengetal2016}, who detected a time delay in back-scattering from the far side of a disk (reverbation).

Light travel time has deeper consequences for resolved imaging. For example, the time for a photon (or, equivalently, a shadow) to travel to a distance $r=100\,$au at the speed of light $c$ is $0.58\,$days. In this amount of time, a clump of disk material orbiting at $0.1\,$au from a $1.5\,$\msun\ star traverses $0.38\,$radians or $6\,$\% of its orbit. Combined with the effect of inclination and a radially varying disk surface height, such a clump can cast shadows with a wide range of curved shapes. In this paper, we model the shape of such features. We note that the same curvature applies to bright features cast by a dip in the height of an otherwise azimuthally uniform rim. For conciseness, we use the term ``shadow'' in the rest of the paper.

In Section~\ref{sec:model}, we present a general model and fitting formula for the of shape of spirals due to light travel time. In Section~\ref{sec:discussion}, we discuss the observational evidence and implications, and present our conclusions in Section~\ref{sec:conclusions}.

\begin{figure}[!ht]
\includegraphics[clip=,width=1.0\linewidth]{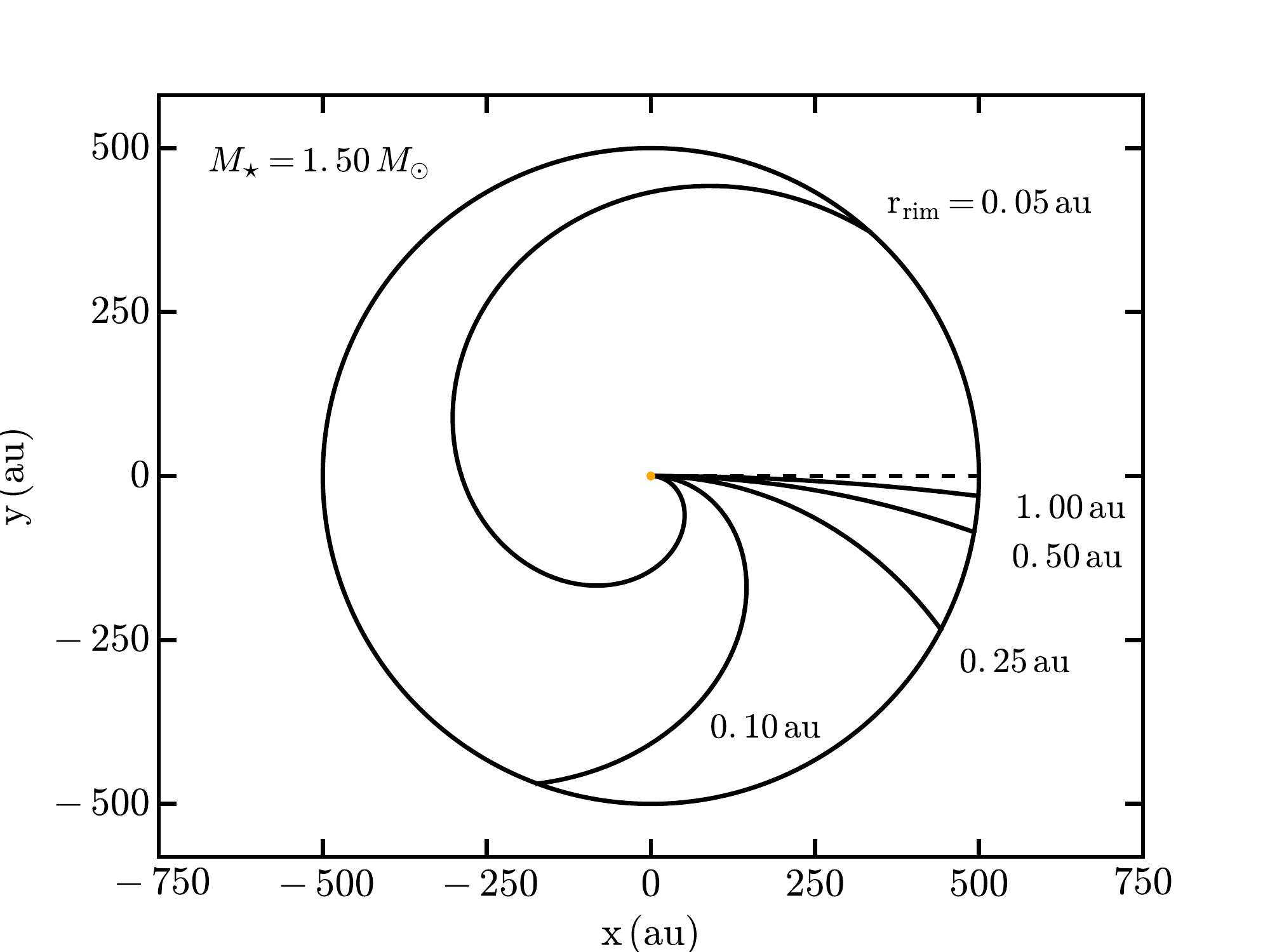}
\caption{Spiral-shaped shadows on a flat, face on disk from light travel time effects. The conical projected shadow (dashed lines) becomes a spiral when finite photon travel times are considered (solid black lines). Spirals from Eq.~\ref{eq:thetalag} are plotted for a $1.5\,M_{\odot}$ star and a range of shadow origin radii, $r_{\rm rim} = 0.05\,$, $0.10$, $0.25$, $0.50$, and $1.00\,$au.}
\label{fig:shadowgallery}
\end{figure}

\section{Spiral features due to photon travel time}\label{sec:model}

We formulate equations for light travel time spirals first for the flat, face-on disk case; and then for the general case of a flared, inclined disk. In discussing the shape of shadows, we consider the set of light rays which simultaneously reach the sky or, equivalently, observer plane. We further assume that all observations are corrected for the disk position angle, such that the semi-minor axis of the disk is vertical, i.e. pointing North.

\subsection{A flat, face-on disk}

The orbital motion of a shadowing clump, located in the inner disk at the radial location $r_{\rm rim}$, induces an angular lag $d\theta_{\rm lag}(r)=\theta_{\rm rim}-\theta_{\rm 1}$ between the clump and a point along the path of the shadow, with midplane radial location $r_{\rm 1}$:
\begin{equation}
d\theta_{\rm lag}(r_{\rm 1}) = \left( \frac{r_{\rm 1}}{c} \right)\, \Omega_{\rm K}(r_{\rm rim}),
\label{eq:thetalag}
\end{equation}

where $\Omega_{\rm K}(r_{\rm rim})=({\rm G}\,M_{\star})^{0.5}\,r_{\rm rim}^{-3/2}$ is the Keplerian angular velocity at the orbital location of the shadowing clump, and $M_{\star}$ is the stellar mass. We define $\theta$ to increase in the direction of orbital motion. The dependence of $d\theta_{\rm lag}$ on $r_{\rm 1}$ describes a spiral, shown in Fig.~\ref{fig:shadowgallery} for several values of the origin radius. As $r_{\rm rim}$ increases, the shadow will become more straight, and will eventually become a straight line for all reasonable disk sizes.

\begin{figure*}[!ht]
\centering
\includegraphics[width=0.90\linewidth]{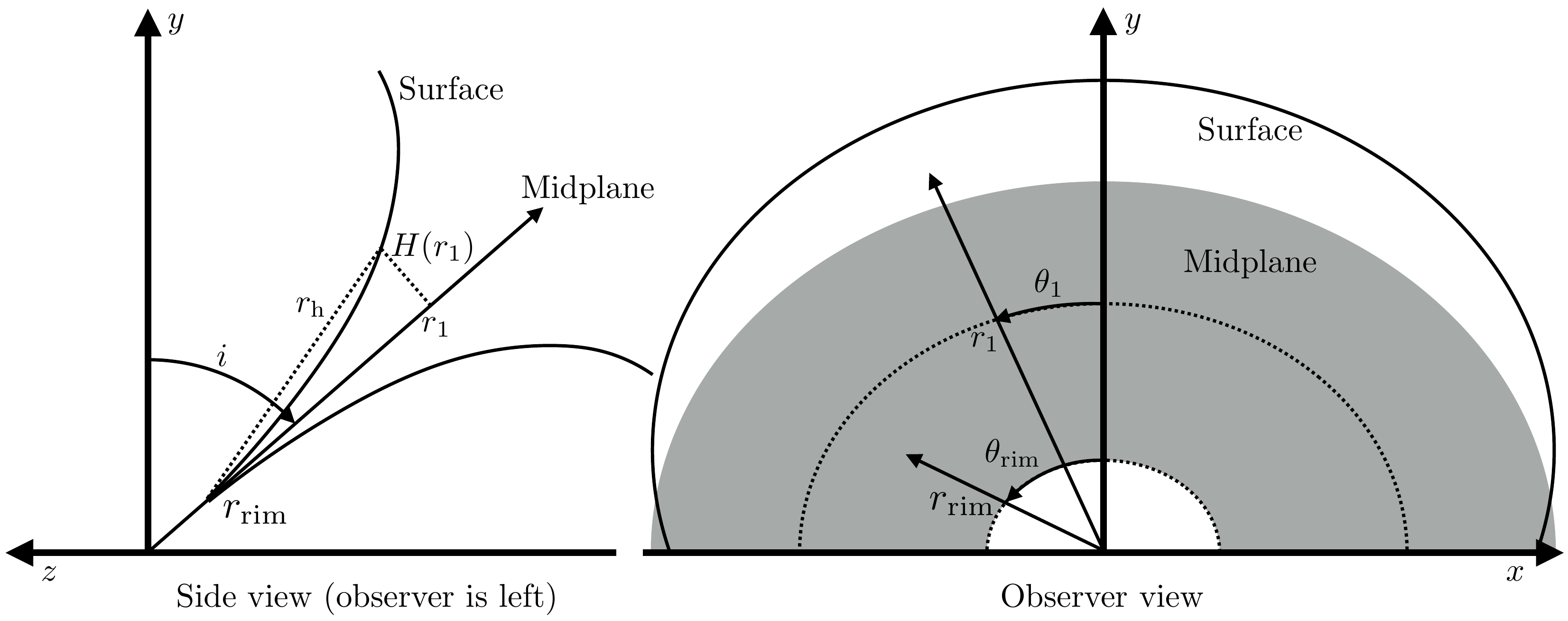}
\caption{Adopted parameterization for a flared and inclined disk.}
\label{fig:geometry}
\end{figure*}

\subsection{A flared, inclined disk}

We now determine the observed shape of a shadow caused by a clump orbiting in the inner part of an arbitrarily flared and inclined disk. Inclination affects the scattered light appearance of pressure wave spirals and disk gaps through blocking and reflection of light, leading to non-standard apparent shapes \citep{Dongetal2016}. It also affects light travel time features, as the paths of reflected rays get substantially lengthened or shortened with changing inclination. The disk scattering surface height $H(r_{\rm 1})$ is taken to be proportional to the gas scaleheight. It generally increases with midplane radius until the very outermost disk, where the surface density drops off and the disk becomes transparent, which we do not include. Adopting the parametric model of \citet{KenyonHartmann1987}, the vertical height from the midplane is
\begin{equation}\label{eq:flaring}
H(r_{\rm 1}) = h_{\rm c}\,\left( \frac{r_{\rm 1}}{r_{\rm c}} \right)^{\psi}\,r_{\rm 1},
\end{equation}

where $h_{\rm c}$ is the opening angle corresponding to $H(r_{\rm c})$ as seen from the star, and $\psi$ is the flaring power. The distance $r_{\rm h}$ which a photon must cross to intersect the disk surface above $r_{\rm 1}$ is
\begin{equation}\label{eq:rh}
r_{\rm h} = r_{\rm 1}\, \left( 1 + \left[ h_{\rm c}\,\left( \frac{r_{\rm 1}}{r_{\rm c}} \right)^{\psi} \right]^{2} \right)^{1/2}
\end{equation}
For a disk seen face-on, the increased travel time to the flared surface increases the curvature of the spiral shadow relative to the flat disk case.

The inclination of a disk adds an extra dimension to light travel time effects, as the light from points along the curve of Eq.~\ref{eq:thetalag} no longer reaches the observer at the same time. In midplane polar coordinates, the spiral is specified by the location of the shadowing clump at the time of observing, $(\theta_{\rm rim}, r_{\rm rim})$, and the midplane location directly under any given point along the observed shadow, $(\theta_{1}, r_{1})$. The geometry is illustrated in Fig.~\ref{fig:geometry}. 

We first consider $t_{\rm r}$, the travel time for a photon from the radial origin of the shadow to a reflecting point on the disk surface, and $t_{\rm z}$, the travel time from that point until a perpendicular intersection with the sky plane at $(z{=}0)$. The observation of a scattered photon following the $t_{\rm z}+t_{\rm r}$ path will coincide with an observed photon from the inner disk, originating from the clump after it spent time $t_{\Omega}$  in Keplerian rotation and having a light travel time $t_{z,{\rm rim}}$ to the sky plane. The four time intervals can be expressed as
\begin{eqnarray}
t_{\Omega} &=& \frac{ \theta_{\rm rim} - \theta_{\rm 1} }{ \Omega_{\rm K} }, \label{eq:tzrim} \\
t_{\rm z,rim} &=& \frac{ r_{\rm rim}\,\cos{(\theta_{\rm rim})}\,\sin{(i)} }{ c },\\
t_{\rm z} &=& \frac{ r_{\rm 1}\,\cos{(\theta_{\rm 1})}\,\sin{(i)} - H_{\rm 1}\,\cos{(i)} }{ c },\\
\textrm{and }t_{\rm r} &=& \frac{ r_{\rm 1} - r_{\rm rim} }{ c }\textrm{, where } r_{\rm rim}\leq r_{\rm 1}.
\end{eqnarray}

In Eq.~\ref{eq:tzrim}, we have assumed that the vertical height of the feature casting the shadow has a negligible impact on the light travel time from $r_{\rm rim}$ to any observable point on the disk surface. Since $t_{\Omega}=d\theta_{\rm lag}/\Omega_{\rm K}$, we can write
\begin{eqnarray}\label{eq:thetalag2}
d\theta_{\rm lag} &=& \theta_{\rm rim} -\theta_{\rm 1} \\
  &=&  \Omega_{\rm K} \times \left( t_z + t_r - t_{z,{\rm rim}} \right).
\end{eqnarray}

Solving Eq.~\ref{eq:thetalag2}, assuming that $\theta_{\rm rim}\geq\theta_{\rm 1}$ and $r_{\rm rim}\leq r_{\rm 1}$, we arrive at
\begin{equation}\label{eq:3dspiral}
r_{\rm 1} = r_{\rm rim}\,\frac{ (\theta_{\rm rim} - \theta_{\rm 1})\,c/v_{\rm K} + 1 + \cos{(\theta_{\rm rim})}\,\sin{(i)} }{ ( 1 + F_{\rm c}^{2} )^{1/2} - F_{\rm c}\,\cos{(i)} + \cos{(\theta_{\rm 1})}\,\sin{(i)}},
\end{equation}

where $F_{\rm c} = h_{\rm c}\,( r_{\rm 1} / r_{\rm c} )^\psi$ parameterizes the vertical structure.

\begin{figure*}[!ht]
\includegraphics[clip=,width=1.0\linewidth]{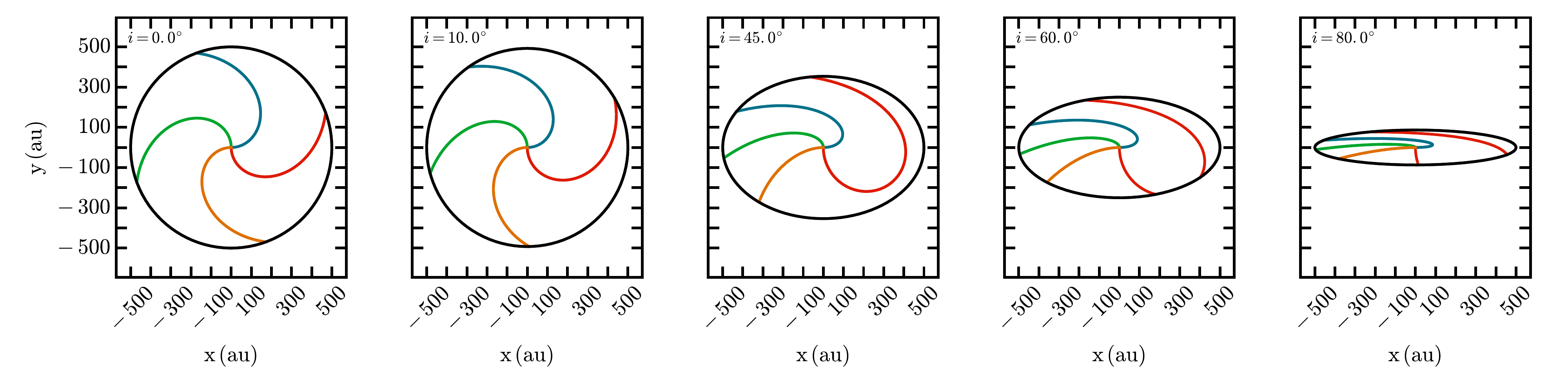}
\caption{Shadows cast on a geometrically flat disk in clockwise rotation and with an outer radius of $400\,$au. The coloured curves indicate shadowing by clumps orbiting at  $r_{\rm rim}=0.1\,$au and with instantaneous orbital phases of $\theta_\textrm{rim}=0$, $\pi/2$, $\pi$, and $3\pi/2\,$rad. The panels, from left to right, show $i=0^{\circ}$, $10^{\circ}$, $45^{\circ}$, $60^{\circ}$, and $80^{\circ}$. }
\label{fig:flat3dinc}
\end{figure*}

In Fig.~\ref{fig:flat3dinc}, we show numerical solutions to Eq.~\ref{eq:3dspiral} for the flat disk case ($h_{\rm c}=0$) with $r_{\rm rim}=0.1\,$au for five different disk inclinations, converted to on-sky $(x,y)$ coordinates according to the equations
\begin{eqnarray}
x &=& r_{\rm 1}\sin{(\theta_{\rm 1})}\\
\textrm{and }y &=& r_{\rm 1}\cos{(\theta_{\rm 1})}\cos{(i)} + r_{\rm 1}F_{\rm c}\sin{(i)}.
\end{eqnarray} 

Viewed face-on, the four spirals originating at points $\theta_{\rm rim}=0$, $\pi/2$, $\pi$ and $3\pi/2$\,rad are all identical. With increasing inclination, their shapes diverge, as the interplay of depth-of-field and Keplerian rotation direction takes effect. The appearance of the spiral feature is radically different when it is on the side of the disk moving towards or away from the observer. For the largest inclinations, the $\theta_{\rm rim}=\pi$\,rad spiral breaks into two arcs as the middle segment of the spiral falls outside of the outer radius of our model disk.

\begin{figure*}[!ht]
\includegraphics[clip=,width=1.0\linewidth]{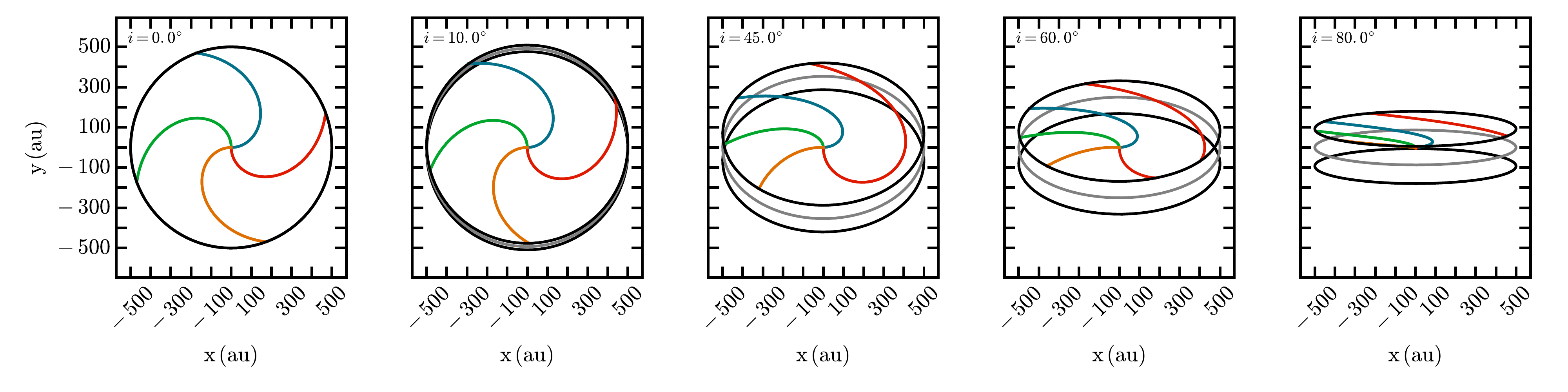}
\caption{Shadows cast on a flared disk in clockwise rotation and with an outer radius of $400\,$au. The coloured curves indicate shadowing by clumps orbiting at  $r_{\rm rim}=0.1\,$au and with instantaneous orbital phases of $\theta_\textrm{rim}=0$, $\pi/2$, $\pi$, and $3\pi/2\,$rad. The panels, from left to right, show $i=0^{\circ}$, $10^{\circ}$, $45^{\circ}$, $60^{\circ}$, and $80^{\circ}$. }
\label{fig:flaring3dinc}
\end{figure*}

In Fig.~\ref{fig:flaring3dinc}, the calculations for the same inner rim and viewing conditions as in Fig.~\ref{fig:flat3dinc} are shown for a flaring disk ($h_{\rm c}=0.1$, $\psi=0.25$, $r_{\rm c}=40\,$au). The three circles correspond to the disk midplane and the forward- and backward-facing outer rim of the flared surface. The spirals are generally more tightly wound than in the flat disk case, and the differences are largest at the highest inclinations. 

In Fig.~\ref{fig:wedges}, the spirals are shown with a finite azimuthal width of $6\,$\% of $2\,\pi$ at their origin ($r_{\rm rim}{=}0.1\,$au), for the case of $i{=}60^{\circ}$, to illustrate the shadow shapes for an obscuring clump of finite azimuthal size. Animations showing the shadows in motion for various inclinations are provided online\footnote{\texttt{dl.dropboxusercontent.com/u/3526708/spiralmovies.zip}}.


\section{Discussion}\label{sec:discussion}

\subsection{Comparison with other spiral-making processes}

Spirals in circumstellar disks may alternatively be formed from pressure waves induced when the disk interacts with itself, an orbiting planetary or stellar companion, or an external envelope; or by magnetohydrodynamical effects at the interface of high- and low-ionisation regions within the disk (a so-called dead zone). In all cases, the time-dependence or  pattern speed of the spiral is simply controlled by the orbital angular-velocity at the spiral origin, proportional to $r^{-3/2}$ where $r$ is the radius of the inner-clump, planet, disk-envelope interface \citep{Lesuretal2015}, or ionisation gradient \citep{Lyraetal2015}.

Giant (proto-)planets capable of generating significant pressure waves are expected to be at radii of $10$s to $100$s of au, and the disk-envelope interface is necessarily at an even larger radius. Dead zones can give rise to azimuthally confined vortices at the inner thermal ionization boundary in a disk, corresponding to gas temperatures of ${\sim}1000\,$K, i.e. radii ${\sim}0.1\,$au \citep{UmebayashiNakano1988, DeschTurner2015}. However, the pressure wave spirals originating from such regions are not sharply defined azimuthally or vertically at $10$s or $100$s of au, appearing as several tighly wound spiral arms \citep{BaruteauLin2010, Lyraetal2015}. The pattern speed of a delayed scattered light spiral caused by a shadowing clump at $0.1\,$au in the inner disk is likely a factor ${\sim}1000$ greater than the likely cases of a detectable pressure wave spiral, in particular due to (sub-)stellar companions or self-gravity.

Spiral curvature is approximately proportional to the ratio of pattern speed and a radial propagation speed. As noted above, the pattern speeds of detectable pressure wave spirals are expected to be significantly lower than for shadow spirals, but their radial propagation, approximately given by the speed of sound (${\sim}100$--$1000\,$m\,s$^{-1}$), is also reduced relative to the speed-of-light propagation for a shadow, by a factor of ${\sim}10^{-6}$. Thus, a light travel time spiral will always be distinct from a pressure wave effect due either its greater pattern speed; lesser curvature; or, more likely, both.

\begin{figure}[!ht]
\includegraphics[clip=,width=1.0\linewidth]{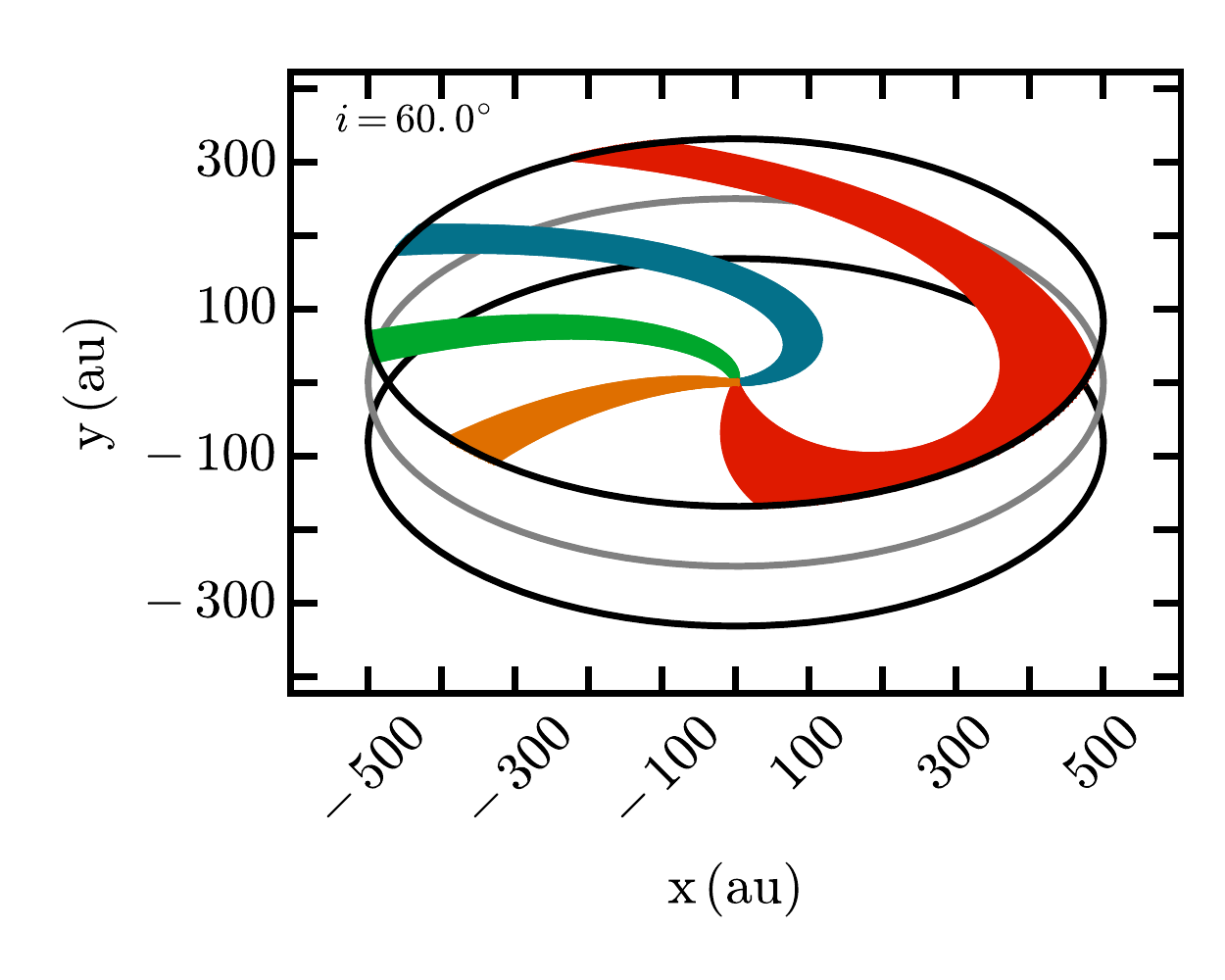}
\caption{Light travel time spirals for an inner disk perturbation shadowing (or, equivalently, illuminating) the disk across $6\,$\% of the azimuthal range. Solutions from Eq.~\ref{eq:3dspiral} are plotted for a $1.5\,M_{\odot}$ star and central origin points at $\theta_\textrm{rim}=0$, $\pi/2$, $\pi$, and $3\pi/2\,$rad. }
\label{fig:wedges}
\end{figure}

An important property of spirals arising from light travel time is that, unlike the limited parameter range of azimuthally-asymmetric features caused by planets, self-gravity, infall envelopes, or Rossby wave instability, a large variety of structures can form, such as single, loosely-wound spiral arms and disjointed arcs, the shapes of which mainly depend on the location of a shadowing clump, disk aspect ratio, and disk inclination. These structures will change rapidly according to the Keplerian frequency of the shadowing clump, and with faster or slower variations in different parts of the disk due to its inclination, as seen in Figs.~\ref{fig:flaring3dinc} and \ref{fig:wedges}.

\subsection{Partial shadowing: umbra and penumbra}

If the physical extent of the shadowing object is smaller than that of the central star, the shadow edge will no longer be well defined. There will be a radially converging full shadow (umbra) and a wider, longer partially shadowed region (penumbra). Our formulation of light travel time spirals assumes that the star is a point source of light, and therefore we cannot currently model such features. Should an umbra and penumbra be observed however, the physical dimensions of the shadowing object relative to those of the star will be strongly constrained, and the centerline of the penumbra will still correspond to Eq.~\ref{eq:3dspiral}.

\subsection{Observability}

Spatially resolved observations of faint and time-variable structures in scattered light have become feasible with recent advances in high-contrast, high angular resolution scattered light imaging, brought about by the SPHERE and GPI instruments in particular. The spiral features predicted by Eq.~\ref{eq:3dspiral} should be observable in such images. During the preparation of this manuscript, \citet{Stolkeretal2016} reported a curved shadow in the HD~135344~B disk and determined that it is well fitted by the flat-disk shadow as described by Eq.~\ref{eq:thetalag}. However, follow-up observations show the feature has not moved on a timescale of ${\sim}1\,$month, inconsistent with the expected fast orbital timescale of a light travel time feature (T.~Stolker, priv.comm.). In Subaru/HiCIAO $H$- and $K$-band scattered light imaging of four outbursting FU~Ori disk systems, \citet{Liuetal2016} displays a $1000\,$au-scale linear feature at an odd angle with respect to the center-radial direction (their Fig.~1). The disk inclination is $i{\approx}60^{\circ}$ \citep{Malbetetal1993} and the observed feature resembles the theoretical red shadows shown in Figs.~\ref{fig:flat3dinc} and \ref{fig:flaring3dinc} for $i{\approx}60^{\circ}$.

In this paper, we have assumed an imageable scattered light disk size of $500\,$au, which is larger than many disks currently accessible to direct imaging. For smaller disks, the full range of features shown in Figs.~\ref{fig:flat3dinc} and \ref{fig:flaring3dinc} will appear if the radial location of the shadowing clump is closer to the star.

Given the rapidly increasing number of resolved, high-contrast disk images at visible and near-infrared wavelengths, and the potential for time-resolved multiple datasets, we are hopeful that the occurrence rate and morphological diversity of light travel time spirals will soon be robustly observationally tested. If they are common, a new tool will be available for constraining the inner and outer disk structure and distance to the systems. If light travel time spirals are not common, constraints can be put on the physical dimensions and survival timescales of major optical depth perturbations in the inner disk.

\section{Conclusions}\label{sec:conclusions}

We have calculated the effect of light travel time on the shape of a shadow or a bright feature cast on a protoplanetary disk by an azimuthal perturbation in the optical thickness of the inner disk for the general case of a flared and inclined disk. We find that such shadows when observed in scattered light can take on a range of curved shapes, in some cases breaking into disjointed and misaligned arcs.

We provide an expression (Eq.~\ref{eq:3dspiral}) describing the shape of light travel time spirals in the general case which allows for easily predicting their form and for fitting of observations.

Fitting the equation to observed shadows allows the constraining of parameters such as the orbital motion of a shadowing clump, the mass of the central star, the outer disk vertical structure and physical scale, and the distance to the system. Very recent scattered-light observations have found structures which resemble these phenomena and may provide a new way to study the structure and dynamics of material in the unresolved inner disk.

\begin{acknowledgements}
The authors thank Tomas Stolker and Wladimir Lyra for useful discussions. This work is supported by a Royal Netherlands Academy of Arts and Sciences (KNAW) professor prize.
\end{acknowledgements}

\bibliographystyle{aa}
\bibliography{shadows}

\end{document}